\documentclass[12pt,aps,tightenlines,endfloats,a4paper]{revtex4}
\usepackage{subfigure}
\usepackage{graphicx}
\begin{document}
\title{Bound State Calculations for Three Atoms Without 
        Explicit Partial Wave Decomposition}
\author{V. A. Roudnev$^{1,2}$, S. L. Yakovlev$^2$, and S. A. Sofianos$^1$}
\address{$^1$ Physics Department, University of South Africa, P.O. 
Box 392, Pretoria 0003, South Africa.
}

\address{$^2$ Department of Mathematics $\&$ Computational Physics,
Institute for Physics St-Petersburg State University
198504 Ul'yanovskaya 1, Petrodvorets, St. Petersburg, Russia.
}

\begin{abstract}
A method to calculate the bound states of three-atoms without resorting to
an explicit partial wave  decomposition  is presented.
The differential form of the Faddeev equations in the  total angular 
momentum representation is used for this purpose. The method utilizes  
Cartesian coordinates combined with the tensor-trick preconditioning 
for large linear systems and Arnoldi's algorithm for eigenanalysis.
As an example, we consider  the  He$_3$ system in  which the interatomic 
force  has a very strong repulsive core that makes the three-body
calculations with standard methods tedious and  cumbersome requiring
the inclusion of a large number of partial waves. The  results
obtained  compare favorably with other results in the field.\\\\
PACS numbers: 21.45.+v, 36.90.+f,02.70.Jn
\end{abstract}
\maketitle
\section{Introduction}
In recent years the  $^4$He trimer has been the center of several 
theoretical investigations (see, for example, Refs.
\cite{Fedorov,Roudnev,Motoold} and references therein). From all methods employed in these studies, 
the Faddeev equations method is perhaps the most attractive since 
it reduces the Schr\"odinger equation
for three particles into a system of integral or differential
equations which can be used to study bound states
and  scattering processes in a rigorous way. 

The differential form of the Faddeev equations has been  proposed  long ago
 by Noyes and Fiedeldey \cite{Noyes}.  Since then, these equations have 
been used in bound states calculations in Nuclear and Coulomb 
systems \cite{GignLav,MerkGignLav,Friar,KviHu,YkFl} 
-- just to mention  a few references.
The derivation  and discussion by  Merkuriev of the asymptotic 
boundary condition  of the  differential Faddeev equations (DFE) 
\cite{Merkuriev}, paved the  way to use the DFE not 
only in bound-state calculations but in investigations of scattering 
processes as well.  For atomic systems, however, some peculiarities 
of the inter-atomic interactions resulted in a limited  use of DFE 
in both   bound and scattering  calculations. 

One of the peculiarities is  that the interaction among atoms often
contains very strong repulsion at short distances which is difficult 
to handle numerically. In addition, it generates 
strong short range correlations implying that 
in atomic systems one should take into account many partial
waves in order to achieve convergence.
These problems can hardly be solved by using the computational
power of modern computers and requires instead the development of
appropriate numerical techniques. 

One such technique is the so-called tensor-trick 
proposed by the Groningen group \cite{Groning, Groning1}. Application of the 
method in Nuclear and Coulomb systems showed that  
high accuracy calculations can be made
using low computer power. 
Another method is that of Cartesian coordinates applied for the first
time in three body Faddeev calculations in \cite{Carbonell}. 
The latter method 
is suitable in describing the long-range behavior of weakly
three-body  bound  states correctly. This feature is crucial in  calculating 
excited states close to the two-body threshold. 
Yet another proposed method of solving the three-body problem is that
of Ref.~\cite{Kvits} in which 
no explicit partial wave decomposition is required and 
only the  total angular momentum of the system is used. Within this
total angular momentum representation method 
the Faddeev operator has a very simple form that allows the
construction of effective numerical schemes.

In the present work we have combined the aforementioned 
methods and, moreover, we  propose a new approach to
overcome the difficulties  arising from the short range
repulsion. We applied the overall  procedure in bound state
calculations of the He$_3$  system using various potentials and
compare our results with other results previously obtained 
by  various groups.

 In Sec. II we describe the equations to be solved and formulate 
the appropriate boundary conditions. In Sec. III  the numerical
method is presented.  Our results are 
given in Sec. IV while our conclusion are summarized in Sec. V. Some
technical details for the field practitioners are shifted in the appendix.
\section{Formalism }
When considering a three-body system it is convenient to denote the two-body
subsystems by  $\alpha $,  $\beta $ and  $\gamma $ and introduce 
Jacobi coordinates  which in the  configuration space 
are defined by
\begin{equation}
\begin{array}{rcl}\cr
           {\bf x}_\alpha&=&\displaystyle\left(\frac{2m_\beta m_\gamma }
		{m_\beta+m_\gamma}\right)^{1/2}
		({\bf r}_\beta-{\bf r}_\gamma)\cr
        {\bf y}_\alpha&=&\displaystyle\left(\frac{2m_\alpha (m_\beta +m_\gamma )}
	     {m_\alpha +m_\beta +m_\gamma }\right )
		\left({\bf r}_\alpha-\displaystyle
               \frac{m_\beta{\bf r}_\beta+m_\gamma
		{\bf r}_\gamma}{m_\beta+m_\gamma }\right)
\end{array}		
\label{Jacoord}
\end{equation}
The coordinates corresponding to other pairs 
 can be obtained by cyclic permutations of the subscripts  $\alpha$,  
$\beta$, and $\gamma$, their relation being
\begin{equation}
\label{JacobTransf}
  \left( 
     \matrix{
	{\bf x}_\beta\cr
	{\bf y}_\beta
            }  \right) 
	=
  \left( 
     \matrix{
	S^{11}_{\beta \alpha } & S^{12}_{\beta \alpha }\cr
	S^{21}_{\beta \alpha } & S^{22}_{\beta \alpha }\cr
     }
  \right) 
  \left( 
     \matrix{
	{\bf x}_\alpha \cr
	{\bf y}_\alpha
                }
  \right) \, 
\end{equation}
where  $S^{ij}_{\beta\alpha}$ are  coefficients which depend on the masses of the
particles \cite{MerkFadd}. For identical particles these coefficients are 
$$
  S^{11}_{\beta \alpha }=  S^{22}_{\beta \alpha } = \frac{1}{2}\,,
  \qquad  S^{12}_{\beta \alpha }= -S^{21}_{\beta \alpha } = \frac{\sqrt{3}}{2}
$$
We assume that the Hamiltonian of the system involves 
only two-body interactions 
\begin{equation}
\label{Hamilt}
	H=H_{0}+\sum _i V_i ({\bf x}_i )\,,
\end{equation}
where  $H_0=-\Delta_x-\Delta_y$ is the Hamiltonian of three free
particles and  $V_i({\bf x}_i)$ are the two-body potentials.\\

The wave function  $\Psi $
of the system can be expressed in terms of the three Faddeev 
components  
$\Phi_i $
\begin{equation}
	\Psi ({\bf x},{\bf y})=\sum _{i }\Phi _{i }
	({\bf x}_{i},{\bf y}_{i})\,
\end{equation}
satisfying the  Faddeev equations
\begin{equation}
\label{FaddEqR6}
 (-\Delta_x-\Delta_y+V_{i}({\bf x}_{i})-E )
	\Phi_{i} ({\bf x}_{i} ,{\bf y}_{i})
	=
  -V_{i} ({\bf x}_{i})\sum _{j \ne i}
	\Phi _{j}({\bf x}_{j},{\bf y}_{j})
\end{equation}
 where  $E$ is the energy of the system. 
In what follows we shall restrict ourselves, without loss of generality,
to  three identical bosons which is the case under consideration. 
In such a case the Faddeev components have the same functional form in
their own coordinates and thus the 
system (\ref{FaddEqR6}) is  reduced to  one equation only
\begin{equation}
\label{FaddEqMatrix}
		(H_0+V-E)\Phi =-V(C^++C^-)\Phi \ .
\end{equation}
In this equation $H_0=-\Delta_x-\Delta_y$, $C^+$ and $C^- $ are  the  cyclic 
and anticyclic permutation
operators respectively, $\Phi$ is one of the Faddeev components
$\Phi_i$ written in its own coordinates.

The potential energy of the system is  invariant with respect to rotations.
This makes it possible to separate out the degrees of freedom corresponding to 
rotations of the system. This can be achieved
 by expanding the Faddeev component  $\Phi$ in terms of eigenfunctions
 of the total angular momentum, {\em i.e.}
Wigner functions $  D_{mn}^{L}(g)$ \cite{Kvits},
\begin{equation}
	\Phi ({\bf x},{\bf y})=
		\sum _{L,m,n}\frac{\phi^{Lmn}(x,y,z)}{xy}\, 
	      D_{mn}^{L}(g)\, .
\end{equation}
Here  $g\in SO(3)$ are the coordinates
describing collective angular motion of the system and  $ \phi ^{Lmn}(x,y,z) $  
are the projections of the Faddeev component in subspaces with fixed angular
momentum. The components of the projections depend on the {\em intrinsic
coordinates}
\begin{equation}
\label{IntrCoord}
	x=|{\bf x}|,\quad y=|{\bf y}|,\quad
		z=\frac{({\bf x},{\bf y})}{xy},\qquad
  		x,y\in [0,\infty ),\ \  z\in (-1,1)
\end{equation}
which describe the intrinsic state of the cluster. Since we can fix the
total angular momentum  $L$ of the system, the corresponding 
projection of the free Hamiltonian can be written as 
\begin{equation}
\label{H0L}
 		H_{0}^{L}=D^{L}(g^{-1})xyH_{0}\frac{1}{xy}D^{L}(g)
\end{equation}
where  $D^{L}(g)$ stands for a matrix constructed from the Wigner functions
\cite{Kvits}. In the case of zero total angular momentum the 
explicit expression for  $H^{0}_{0}$ reads 
\begin{equation}
\label{H00}
	H^{0}_{0}=-\frac{\partial ^{2}}
	{\partial x^{2}}-\frac{\partial ^{2}}
	{\partial y^{2}}-(\frac{1}{x^{2}}
	+\frac{1}{y^{2}})\frac{\partial }
	{\partial z}(1-z^{2})
	\frac{\partial }{\partial z}\,.
\end{equation}
The corresponding projection of Eq. (\ref{FaddEqMatrix}) takes the
form
\begin{equation}
\label{EqFaddTAM1}
\displaystyle       
   (H^0_0+V(x)-E)\phi^0(x,y,z)
   =
   -V(x) P \phi^0(x,y,z) \,,
\end{equation}
where 
\[P \phi^0(x,y,z) 
      \equiv 
      xy(
        \frac{\phi^0(x^+,y^+,z^+)}{x^+ y^+}
	+\frac{\phi^0(x^-,y^-,z^-))}{x^- y^-}
	)
\]
and $x^\pm(x,y,z)$, $y^\pm(x,y,z)$, and $z^\pm(x,y,z)$ are
\begin{equation}
\label{xyzpm}
  \begin{array}{rcl} \displaystyle
     x^\pm (x,y,z)& = & \displaystyle 
                        (\frac{1}{4}x^2
                              +
			 \frac{3}{4}y^2
			      \mp
			 \frac{\sqrt{3}}{2}xyz
			 )^{1/2} 
			 \; ,    \\  
     y^\pm (x,y,z)& = & \displaystyle
                        (\frac{3}{4}x^2
                              +
			 \frac{1}{4}y^2
			       \pm
                         \frac{\sqrt{3}}{2}xyz
			 )^{1/2}
			 \; ,   \\ 
     z^\pm (x,y,z)& = & \displaystyle
                        \frac{\displaystyle\pm \frac{\sqrt{3}}{4}x^2
                              \mp \frac{\sqrt{3}}{4}y^2
		               - \displaystyle \frac{1}{2}xyz}
			     {x^{\pm }(x,y,z)\, y^{\pm }(x,y,z)}\; .
  \end{array}
\end{equation}
Assuming that in each two-body subsystem only one bound
state  exists, we can write the asymptotic boundary conditions 
for the Faddeev component  $\phi^0$ as follows
\begin{equation}
\label{Asym}
	\phi^0(x,y,z)\sim \, \varphi_2(x)\,
	{\rm e}^{-k_y y}+A(x/y,z)\,
	\frac{{\rm e}^{-k_3(x^2+y^2)^{1/2}}}{(x^{2}+y^{2})^{1/4}}\,,
\end{equation}
where  $\varphi _{2}(x)$ denotes the wave function of the two-body bound
state in the two-body subsystem,  $k_{y}=\sqrt{E_{2}-E_{3}}$,  
$k_{3}=\sqrt{-E_{3}}$,  $E_{2}$ is the two--body bound state energy,
and  $E_{3}$ the energy of the three-body  system. 
The first term corresponds to virtual decay into a particle 
and a two-body bound system,
usually denoted as 2+1, while the second term corresponds to a virtual decay
with an amplitude $A(x/y,z)$ into  three single particles denoted as  1+1+1. 
The term corresponding to the latter configuration 
decreases much faster than in the 2+1.\\

In the present work the term corresponding to 1+1+1 virtual 
decay is neglected  and thus the asymptotic  boundary conditions 
for the Faddeev component at sufficiently
large distances  $R_x$ and  $R_y$  read
\begin{equation}
\label{ApprBC}
	\left .\frac{\partial }{\partial x}\ln\phi ^{0}(x,y,z)
		\right |_{x=R_x}=-k_{x}\equiv i\sqrt{E_2}\,,\qquad
	\left .\frac{\partial }{\partial y}\ln\phi ^{0}(x,y,z)
          \right |_{y=R_y}=-k_y\,.
\end{equation}
\section{Numerical Procedure}
The first  numerical calculation with the DFE were made by
Laverne and Gignoux in the early seventies \cite{GignLav}. Since then,
many numerical methods were proposed. 
In the present work we shall employ: i) Quintique splines
\cite{KviHu} together with an  orthogonal collocation \cite{deBoorSwartz} 
procedure which allows us  to construct a linear system 
of equations corresponding to  Eq. (\ref{EqFaddTAM1}). 
In this way the bound state problem can be transformed 
to a  generalized eigenvalue problem. ii) The tensor-trick 
 \cite{Groning} to solve the eigenvalue problem using the  restarting 
Arnoldi algorithm \cite{Saad}.

Any regular solution of Eq. (\ref{EqFaddTAM1}) fulfilling
the boundary  conditions (\ref{ApprBC})
 can be approximated by an  expansion in terms of the basic functions
(see Appendix A)
\begin{equation}
\label{FadCompSplinExpa}
	\phi^0(x,y,z)=\sum_i f_i B_{i}(x,y,z)
\end{equation}
where   $i$ stands for a multi--index  $\{i_{x},i_{y},i_{z}\}$
and  $B_{i}(x,y,z)=B_{i_{x}}(x)B_{i_{y}}(y)B_{i_{z}}(z)$. 
Following the procedure described in the Appendix A we obtain an equation for the
coefficients $f_i$:
\begin{equation}
\label{discrFadd}
           (\hat H_F-E\hat I){\bf f}=0 \, ,
\end{equation}
where $\hat I$ is the unit matrix and $\hat H_F$ is the discrete 
analog of the Faddeev operator (see Appendix A)
\begin{equation}
\label{HF}
	 \hat H_F= \hat H^0_0 + \hat V (\hat I + \hat P ) \, .
\end{equation}
When investigating nuclear systems with short-range potentials,
such as the Malfliet-Tjon V (MT-V) potential \cite{mt},
the  spectrum of the system (\ref{discrFadd}) can be  calculated
by direct application of the restarting Arnoldi or biorthogonal Lanczos
algorithm. However, an additional regularization is usually needed
to accelerate the convergence. For this, one may split the operator $\hat H_F$
in Eq. (\ref{discrFadd}) into two parts,  
$ \hat H_F=\hat H_1+\hat H_2 $,  where $ \hat H_1$,
is such that the operator  $(\hat H_1-E\hat I)^{-1}$  
can be explicitly constructed while $ \hat H_2$ can be considered 
as a remainder.
The technique of explicit inversion of an operator  $\hat H_1-E\hat I$  
is known as tensor-trick \cite{Groning} and is briefly described 
in Appendix B for Cartesian coordinates. \\

Following the standard procedure  we  write for the
eigenvalue problem
\begin{equation}
\label{Lambda1}
	-(\hat H_1-E\hat I)^{-1}\hat H_2{\bf f}=\lambda (E){\bf f}\: .
\end{equation}
where for physical solutions we have  $\lambda (E)=1 $. 
Eq. (\ref{Lambda1}) can be solved using  the Arnoldi algorithm
\cite{Saad}.\\

Although the above  scheme can be applied effectively  
for a wide range of potentials, it is not satisfactory
for interatomic potentials with  a strong repulsive core.
A typical example is the helium dimer  potential which has a
core with  enormously strong repulsion and an extremely weak 
attractive well just enough to hold a two-body bound state. 
In what follows  we give a brief analysis of the 
difficulties that arise and propose a recipe to overcome these 
difficulties.

\begin{figure}
\begin{center}
  \subfigure[TTYPT potential]{
    \includegraphics[width=0.75\textwidth]{TTYDemo_.epsi}
  }
  \subfigure[MT-V potential]{
   \includegraphics[width=0.75\textwidth]{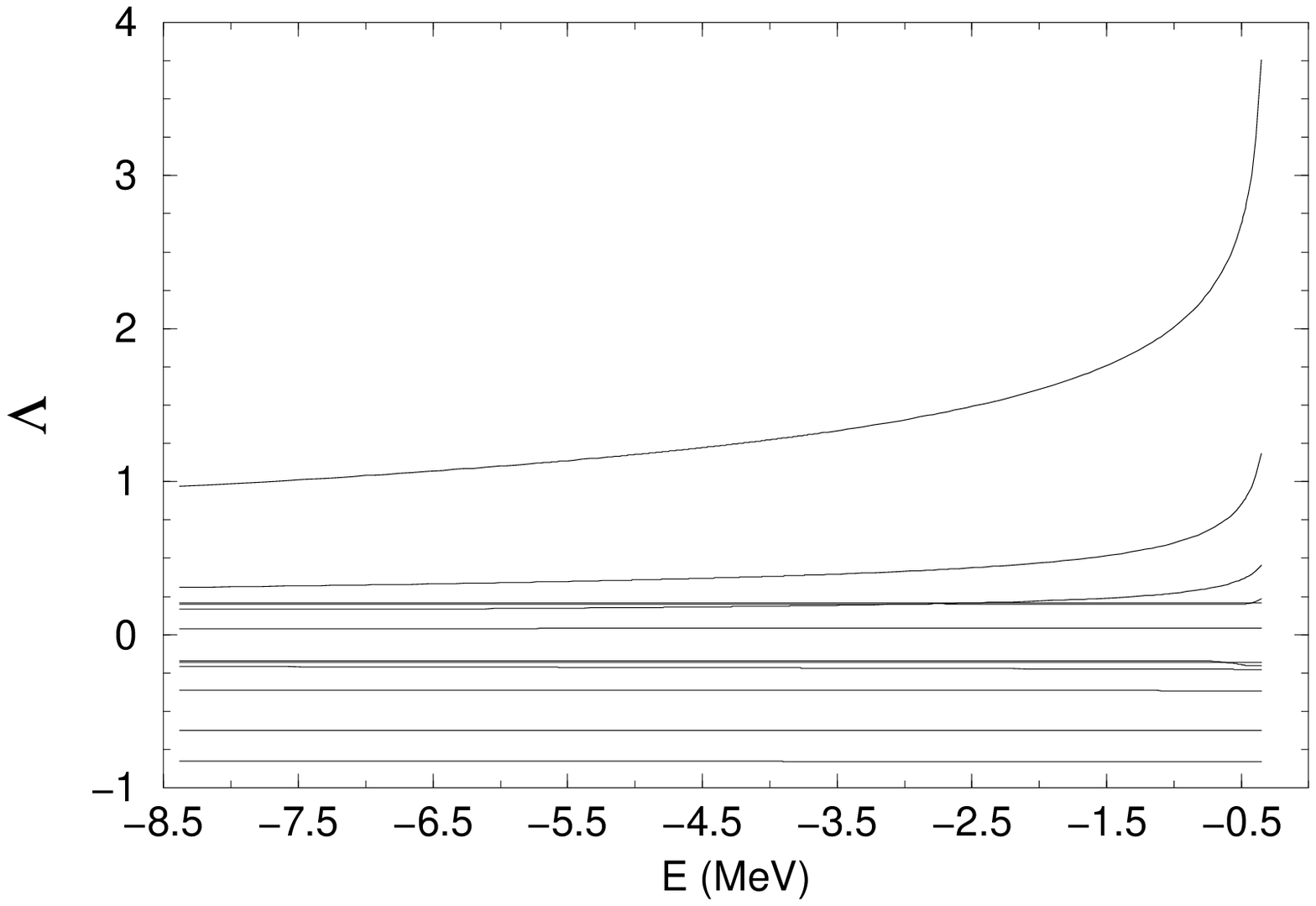} 
  }
  \parbox{0.7\textwidth}{
    \caption{Spectra of operators \(\hat{L}(E)\) for potentials with (a) 
            and without (b) a strong repulsive core 
	    \label{spectra} }
  }
\end{center}
\end{figure}

Since to reproduce the correct long-range behavior of the Faddeev 
components  Cartesian
coordinates are used, the natural choice of the operator $\hat H_1$ in
(\ref{Lambda1}) is 
$$
  	\hat H_1=\hat H_0^0+\hat V \: .
$$
In this case the problem of calculating  bound states is reduced 
to calculation of the largest eigenvalues of the operator
\begin{equation}
\label{Lz}
  	\hat L(z) \equiv -(\hat{H}_{0}+\hat{V}-z)^{-1}
              \hat{V}(\hat{C}^{+}+\hat{C}^{-})
\end{equation}
To demonstrate how the presence of a short-range repulsion manifests 
itself in the spectral properties of the operators  $ \hat L(z) $, one must
compare the spectra of  potentials with and without a
strong repulsive core. Two such potentials are the interatomic He--He
 TTYPT potential \cite{TTY} and the nucleon-nucleon  (MT-V) potential \cite{mt}
the spectra of which are shown in  Fig.~\ref{spectra}. Two main features of the 
spectrum of the TTYPT potential should be pointed out, namely, that  the spectrum
contains numerous  large  negative eigenvalues
and that for all values of the energy $ z $, $ \hat L(z) $  has a number
of  eigenvalues close to one.
According to the estimations of the Arnoldi algorithm convergence rate
(see Appendix C), these are the features that make the convergence 
unacceptably slow. \\

In order to improve convergence we  first look for the origin of the 
negative part of the spectrum. For this we consider the values $z_i$
such as $ \Lambda_i(z_i)\equiv -1$  of the operator  $ \hat L(z) $.
Evidently, they are simultaneously  eigenvalues  of the operator 
$$
	 \hat H_{\rm negative} \equiv \hat H_0+\hat V
                         -\hat V \hat P \, .
$$
Keeping in mind that the operator  $ \hat V $  stands for a 
potential with extremely strong short-range repulsion, we conclude that the
term $ -\hat V \hat P $  
produces extremely strong
short-range attraction. Therefore, the operator  $ \hat H_{\rm negative}$  
contains  a large number of  eigenvalues in the range  $ [-\infty, E_2] $
and their presence suppresses the convergence of the
Arnoldi algorithm for values close to one. 

To eliminate the negative eigenvalues from the spectrum,
 we introduce, instead of $\hat L(z)$, the operator
\begin{equation}
  	\hat M(z) \equiv (\hat H_0+\hat V+\hat V_m-z)^{-1}
              \left(\hat V_m-\hat V \hat P \right) \, ,
\label{Mz}
\end{equation}
where  $ \hat V_m $  corresponds to  some strong
short-range repulsive potential. It will be referred to as modifying potential.
It can be shown  that the eigenvalues $ \mu_i(z_i)=1 $ of the
operator  $ \hat M(z) $ correspond to the eigenvalues $z_i$ 
of the original Faddeev operator and this does not depend upon the modifying potential. 
In contrast, the eigenvalues   $ \mu_i(z)=-1 $  correspond to the eigenvalues
 $ z_i $  of the equation 
\begin{equation}
	\left[\hat H_0+2\hat V_m+\hat V (1 -\hat P)
                  -z_i\right] {\bf f}=0 \, .
\label{Mzneg}
\end{equation}
Obviously, these eigenvalues depend on the choice of the modifying potential
 $\hat V_m $  a proper choice of which
will eliminate all  eigenvalues of the equation (\ref{Mzneg}) in the range  
$ z_i \in (-\infty, E_2) $. Thus, according to Arnoldi algorithm convergence rate 
estimations, this feature of  $ \hat M(z) $  should dramatically improve 
the convergence to the maximal eigenvalue close to one. \\

The convergence rate can be  further improved if instead of the
operator $ \hat M $  
a properly selected function  of $ \hat M $  is used in the calculations. 
One  such function  can be an  even power of  $ \hat M(z) $  or the 
normalized Chebyshev polynomial of the second kind  $ U_k(\hat M(z))/U_k(1) $. 
It allows an upward shifting  of  the lower bound of the spectrum 
and the  increase of 
the separation between the physically interesting eigenvalues
close to one and the rest part of the spectrum. According to the 
convergence rate estimations these features allow the reduction of the 
computation time and memory requirements considerably.

\section{Results}
We applied the aforementioned numerical procedure to study the 
spectrum of the  He$_3$ trimer system. For this purpose, we employed
the most recent  He--He potentials, namely, the   
SAPT1, SAPT2, and HFD-B3-FC11 potentials \cite{Az97}, and compare the 
results with  those previously obtained \cite{Roudnev}
 using the  LM2M2 \cite{Az91} and  TTYPT \cite{TTY} 
interactions, as well as with those  obtained via other methods.\\

In  Table~\ref{tabTrimerEv} we present the results for the
trimer ground and excited states. We also present, in the same table, 
the results obtained for the  dimer binding energy  $ E_2 $
calculated on the same grid used in three-body calculations as well as the 
exact results for the dimer $ E_2^{\rm ex} $. The  difference 
between these values can be regarded as the lower bound for the 
error of our approach. \\

In  Tables~\ref{tabConv2} and~\ref{tabConv}  we demonstrate 
the convergence, for the LM2M2 potential, of the calculated energies 
with respect to the number of grid points used. It is seen that
convergence is achieved with a relatively low number of the $z$ points
($\sim 12$ points) while for the other coordinates the number
is much larger ($\sim100 $ points). \\

Our results for the LM2M2 potential are given in Table~\ref{tabCompare}
together  with those  obtained by Motovilov {\em et. al.} 
\cite{Moto1,Moto2} via  Faddeev-type equations (boundary condition 
model (BCM)) and by Nielsen {\em et al.}    \cite{Fedorov}
via  the hyperspherical adiabatic approach.
It is seen that an overall good agreement is achieved with
these approaches. However,  for the  ground state a slightly better  agreement
is  achieved with the results  of Ref. \cite{Moto2}  while
for the excited state is with those of Ref. \cite{Fedorov}. 
Comparing our results with the
results of \cite{Moto2} we should emphasize, that the angular basis
used by Motovilov {\em et. al.} \cite{Moto2} coincide with our spline 
basis for the simplest grid that can be used in the present method.
 Performing calculations with this simplified grid, we recovered 
all the digits of the ground state energy reported in \cite{Moto2}. This
confirms the accuracy of their result  and the suitability of the
present method for bound state calculations.  The
better agreement for  the excited state with the one obtained in 
Ref.~\cite{Fedorov} can be attributed to the 2+1 contribution 
to the  Faddeev component for the excited states. 
Taking into account this term in  the BCM is very difficult, 
whereas  the hyperspherical adiabatic approach of Nielsen {\em et. al.}
 \cite{Fedorov}  is more suitable for this purpose.
However, the ground state energy of \cite{Fedorov}  is about
1\% less than our result. Consideration of the geometric properties of
helium trimer can clarify the possible nature of the latter difference.\\

The characteristic size of the bound states of the trimer can be
estimated  either  by calculating  $ \langle r \rangle $  or 
 $ \langle r^{2}\rangle^{1/2} $. The results obtained
for these radii are presented in Tables~\ref{tabR} and~\ref{tabR2}.
It is seen  that they are  approximately 10
 times less  than the radii  of the  dimer molecule. However, the size
of the excited state has the same order of magnitude as that
of the  dimer.\\

\begin{figure}
\begin{center}
  \includegraphics[width=0.75\textwidth]{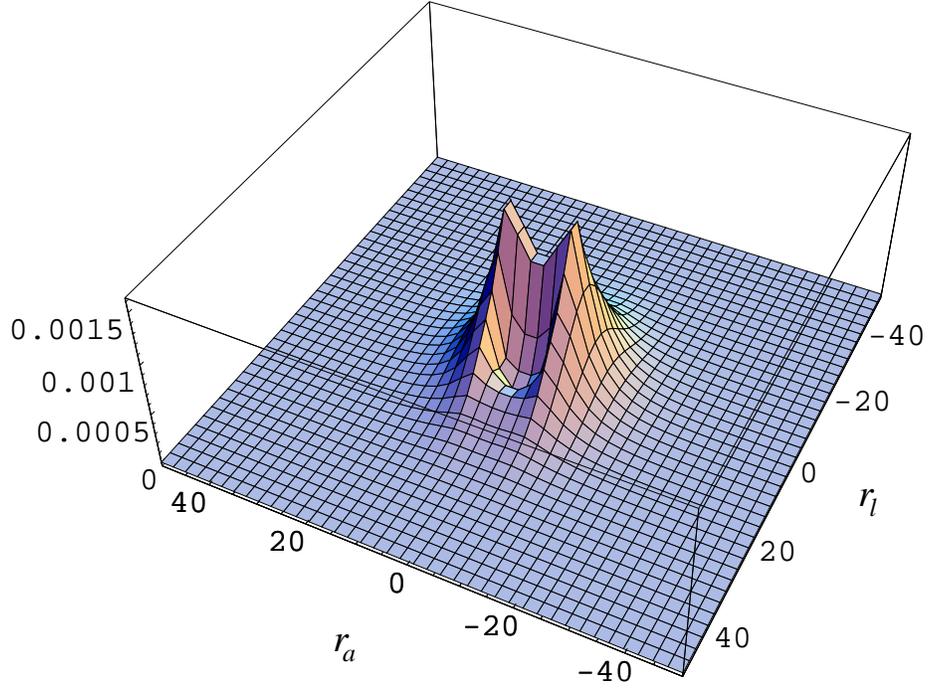}
  \parbox{0.7\textwidth}{
    \caption{ Conditional one-particle density function 
         of the He$_3$ ground state, $r_l$ and $r_a$ are given in \AA
         \label{LM2M2G}}
   }
\end{center}
\end{figure}

\begin{figure}
\begin{center}
  \includegraphics[width=0.75\textwidth]{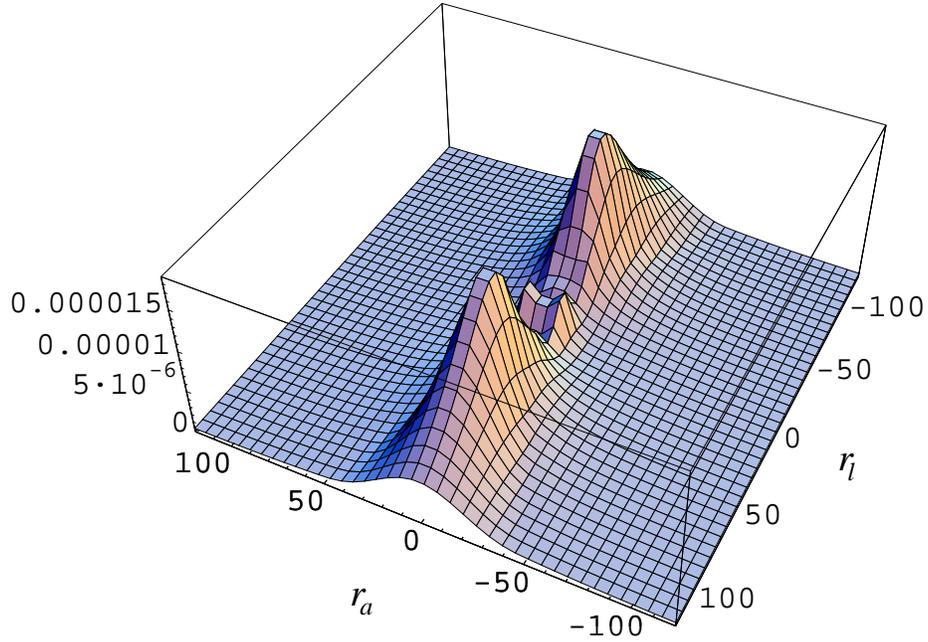} 
  \parbox{0.7\textwidth}{
    \caption{ Conditional one-particle density function 
          of the He$_3$ excited state, $r_l$ and $r_a$ are given in \AA
          \label{LM2M2Ex}} 
  }
\end{center}
\end{figure}

The wave function of the system can be easily obtained from the
Faddeev component,
$$
	  \psi (x,y,z)=\phi (x,y,z)+xy\, \left(\frac{\phi (x^+,y^+,z^+)}
	{x^+y^+}+\frac{\phi (x^-,y^-,z^-)}{x^-y^-}\right) \, ,
$$
where  $ x^\pm$, $ y^\pm$  and  $ z^\pm$ are defined by (\ref{xyzpm}). 
The most intuitive way to visualize the results of the calculations
is to draw the  one-particle density function defined as
$$
  	\rho ({\bf r}_1)=\int \,{\rm d}{\bf r}_2{\rm d}{\bf r}_3 |
		\,\Psi ({\bf r}_1,{\bf r}_2,{\bf r}_3)|^2\, ,
$$
where 
$$
	\Psi({\bf r}_1,{\bf r}_2,{\bf r}_3)=
	   \frac{\psi(x({\bf r}_1,{\bf r}_2,{\bf r}_3),
	  y({\bf r}_1,{\bf r}_2,{\bf r}_3),z({\bf r}_1,
	   {\bf r}_2,{\bf r}_3))}{2\pi x(
	   {\bf  r}_1,{\bf r}_2,{\bf r}_3) y(
	   {\bf r}_1,{\bf  r}_2,{\bf r}_3)} \, .
$$
The functions  
	$ x({\bf r}_1,{\bf r}_2,{\bf  r}_3)$,
        $ y({\bf r}_1,{\bf r}_2,{\bf r}_3)$,  and 
        $ z({\bf r}_1,{\bf r}_2,{\bf r}_3) $  
are obtained from the Jacobi coordinates (\ref{Jacoord}) and (\ref{IntrCoord})
and the wave  function  $ \psi(x,y,z) $  is normalized to unity.
Due to the symmetry properties of the system, the one-particle 
density function depends only on the  coordinate  $ r=|{\bf r}_1| $.
Taking into account the relation  $ {\bf y}_1=\sqrt{3}{\bf r}_1 $  
we get
$$
       \rho (r)= \frac{\sqrt{3}}{4 \pi^2 r^2}
              \int\, {\rm d}x\, {\rm d}z\, | \psi (x,r \sqrt{3},z)|^2\: .
$$
Omitting the integration over  $z$,  we obtain the conditional 
density function $ \rho (r,z) $  describing a spatial distribution 
for particle one when the other two particles are located along a
fixed axis. It is useful to plot this function in coordinates  
$ (r_l,r_a) $  such that  $ r_l=r z $  is
a projection of the position of  particle 1 onto the axis connecting 
the other particles and 
$$
            r_a=\frac{z}{|z|} r(1-z^2)^\frac{1}{2} 
$$
is a projection to the orthogonal 
axis. Three-dimensional plots of the function  
$ \rho ((r_l^2+r_a^2)^{1/2},\cos \arctan r_l/r_a) $  
corresponding to the ground and excited
states of the trimer calculated with the LM2M2 potential are presented in
Figs.~\ref{LM2M2G} and \ref{LM2M2Ex} respectively. The conditional density 
function of the ground state decreases in all directions in a similar way. 
The density function of the excited state has two distinguishable maxima
 and  exhibits  the linear structure of the cluster. This structure 
has a simple physical explanation:  The most probable positions
of a particle in the excited state lie around the other two particles
and when the latter particles are well separated the third one forms 
a dimer-like bound state with each of them. This interpretation agrees with the 
clusterisation coefficients  presented in the Table~\ref{tabClusterisat}. 
These coefficients are calculated as a norm of the function  
$ f_c $  defined by 
\begin{equation}
     	f_c(y)=\int \,{\rm d}x{\rm d}z\, \phi (x,y,z)\varphi_2(x)\,  ,
  \label{ClusterContrib}  
\end{equation}
where  $ \varphi_2(x) $  is the dimer wave function. The values of  
$|f_c(y)|^2$, shown in the Table~\ref{tabClusterisat},
demonstrate the dominating role of a two-body contribution
to the trimer excited state  whereas in the ground state this contribution
is rather small. We could suppose that this dominating contribution of the cluster
wave in the excited state  ensured the fast convergence of the hyperspherical
adiabatic expansion \cite{Fedorov} to the correct value, but in order  to
get the same order of accuracy for the ground state,  possibly more basic 
functions should be taken into account. \\

The advantage  of using  Faddeev equations over 
the Schr\"odinger one in bound-state calculations, is deduced
from the results shown  in Tables~\ref{tabAngContribFC} 
and \ref{tabAngContribWF}.
In the latter table
we  present the  contribution of different angular states to the 
Faddeev component and to the wave function calculated as
\begin{eqnarray}
\nonumber
  	C_n &=& |f_{n}(x,y) |^2 \\ 
  	f_n(x,y)&=&\int^1_{-1}{\rm d}z\, F(x,y,z)P_n(z)\, ,
\label{angularcontr}
\end{eqnarray}
where  $ P_n(z) $, $ n=0,2,4,\cdots $,  are the  Legendre polynomials and
$ F(x,y,z) $ is the Faddeev component or the wave function.
 The angular coefficients
for the Faddeev component decrease much faster than the wave function coefficients.
This could explain the difference between the estimations of the
trimer mean square radius reported in \cite{Roudnev} and the one
presented here. In the previous paper the same angular basis for 
the Faddeev component and the wave function  has been used.
However, it turned out that the
contribution of the higher partial waves to the wave function is not
negligible. This  leads, for instance, to the change of the
 mean square radius of the excited state for the 
 LM2M2 potential from 60.85~\AA \ to
59.3~\AA. This observation agrees with the results of Nielsen
{\em et al.} \cite{Fedorov}, who reported the value 60.86~\AA \
 using essentially the same angular basis both for the wave function 
and for the Faddeev component.
The Table~\ref{tabAngContribFC} also demonstrates that more angular functions
should be taken into account in the ground state calculations.
\section{Conclusions}
 We present  a method which can be used to perform accurate three-body bound
states calculations. It is specially suited for  systems
where the interparticle forces have a strong repulsive core. 
The method is based  on  Faddeev equations  in the   total angular 
momentum representation and without any
further partial wave decomposition. The equations are expressed 
in terms of Cartesian coordinates which are best suited to
describe the long-range behavior of weakly
three-body  bound  states.  Combined with the tensor-trick preconditioning 
for large linear systems and the Arnoldi's algorithm for eigenanalysis
it provided  us accurate results for the the He$_3$ trimer bound and excited
states.\\

 Results obtained with the most recent  realistic intermolecular forces 
 (SAPT1, SAPT2, HFD-B3-FC11), as well as with earlier forces (LM2M2,
TTYPT)  indicate that only two bound states exist. The properties of 
these states are very different with the ground state being  strongly
bound while the binding energy for the  excited state is comparable 
to that of the dimer. The latter implies that a dimmer  cluster
within the molecule is well formed. \\

 The sizes of these two states also differs much. The characteristic
size of the ground state either estimated by  $ \langle r \rangle $  or 
 $ \langle r^{2}\rangle^{1/2} $  is $\sim10$
 times less  than the size of dimer molecule, but the size
of the excited state has the same order of magnitude with that of the dimer.
 This estimation make it  necessary to check for the absence of 
trimers in the experimental media during the measurement of dimer 
properties and vice versa. 
\appendix 
\section{Reduction to a Linear System}
Suppose  $\tau _{t}=\{t_{0},t_{1},\ldots ,t_{N_{t}}\}$ is a partition in 
the range  $[t_{0},t_{N_{t}}]$ in coordinate  $t$ and let  
$S_{5,2}(\tau _{t})$   be the space of quintique Hermite splines 
associated with this partition {\em i.e.}
 piecewise fifth order polynomial functions having two continuous derivatives
in this  range. The partition  $\tau _{t}$ will be
refereed to as the base grid. The set of basic functions in the space  
$S_{5,2}(\tau _{t})$  can be defined by the following conditions
\begin{eqnarray}
\nonumber
&&	S_{i}^{1}(t_{i})=1\,, \qquad 
	\partial _{t}S_{i}^{1}(t_{i})=0\,,\qquad
	 \partial ^{2}_{t}S_{i}^{1}(t_{i})=0\\
\label{Basis0}
&&
	S_{i}^{2}(t_{i})=0\,, \qquad 
	\partial _{t}S_{i}^{2}(t_{i})=1\,,\qquad  
	\partial ^{2}_{t}S_{i}^{2}(t_{i})=0\\
\nonumber&&
	S_{i}^{3}(t_{i})=0\,,\qquad  
	\partial _{t}S_{i}^{3}(t_{i})=0\,,\qquad
	\partial ^{2}_{t}S_{i}^{3}(t_{i})=1
\end{eqnarray}
$ \forall t\notin [t_{\max (0,i-1),},t_{\min (i+1,N_{t})}]\: 
\Rightarrow \: S^{j}_{i}(t)=0 $.
 
For simplicity we omit the subscript  $\tau _{N_{t}}$  
corresponding to any particular choice of a grid in each coordinate. Let the
basic functions be arranged as follows:
\begin{equation}
\label{SplBasIni}
	\widehat{B}_{i}(t)=S^{j}_{k}(t)\,,
	\quad  k=0\ldots N_{t}\,,\ \ j=1\ldots 3\,,
	\ \  i=3k+j-1
\end{equation}
The solution of the Eq. (\ref{EqFaddTAM1}) must satisfy Eq.  (\ref{ApprBC})
and vanish on the planes  $x=0$ and  $y=0$. 
To achieve this
we modify the set of basic functions (\ref{Basis0}) by constructing 
linear combinations  $B_{i}(t)$ which for  $t=x,y$  read
\begin{eqnarray}
\nonumber     &&	
        B_{i}(t)\equiv \widehat{B}_{i}(t)\,,\qquad  i=1\ldots 3N_{t}-1\\
&&
\label{SplineBases}\\
&&\nonumber
	B_{3N_t}(t)\equiv \widehat{B}_{3N_t}(t)-k_t
	\widehat{B}_{3N_t+1}(t)+k^2_{t}\widehat{B}_{3N_t+2}(t) 
\end{eqnarray}
and for the coordinate  $z$ they coincide with the initial basic functions
(\ref{SplBasIni})   $B_{i}(z)=\widehat{B}_{i}(z)$. We look for an approximate
solution of Eq. (\ref{EqFaddTAM1}) in the form of expansion in terms
of the basic functions (\ref{SplineBases})
\begin{equation}
\label{FadCompSplinExp}
	\phi^0(x,y,z)=\sum _{i}f_iB_i(x,y,z)
\end{equation}
where  $i$ stands for the  multi--index  $\{i_{x},i_{y},i_{z}\}$ and  
$B_{i}(x,y,z)=B_{i_{x}}(x)B_{i_{y}}(y)B_{i_{z}}(z)$.
Since the basic functions (\ref{SplineBases}) fulfill the  boundary 
conditions (\ref{ApprBC})
these conditions are also satisfied by the approximate 
component  $\phi^0$.

To obtain a system of equations for the coefficients  $f_i$ the expansion
(\ref{FadCompSplinExp}) should be substituted into Eq. (\ref{EqFaddTAM1})
with some subsequent projection procedure. A reasonable choice of such a
procedure is the method of orthogonal collocations which do not
 require any integration.
The highest order of approximation is achieved by a careful choice
of collocation points for a given basic grid  $\tau $ \cite{deBoorSwartz}.
For the grids  $\tau _{N_x}$,  $\tau _{N_y}$ and  $\tau _{N_z}$  
one should construct the grids of collocation points  
$\tau _{N^{c}_{x}}=\{x^{c}_{1},x^{c}_{2},\ldots ,x^{c}_{N^{c}_{x}}\}$,
 $ \tau _{N^{c}_{y}}=\{y^{c}_{1},y^{c}_{2},\ldots ,y^{c}_{N^{c}_{x}}\} $, and
 $ \tau _{N^{c}_{z}}=\{z^{c}_{1},z^{c}_{2},\ldots ,z^{c}_{N^{c}_{x}}\} $ 
containing
 $ N^{c}_{x}=3N_{x} $, $ N^{c}_{y}=3N^{c}_{y} $, and  $ N^{c}_{z}=3(N_{z}+1) $  
points respectively. Given these collocation grids
the matrix elements of the operators involved in Eq. (\ref{EqFaddTAM1})
can be easily constructed. For example, the matrix elements of the 
identity operator are  $ [I]_{ij}=B_i(x^c_{j_x},y^c_{j_y},z^c_{j_z})$ with  
$ i=\{i_x,i_y,i_z\} $  and  $ j=\{j_x,j_y,j_z\} $. 

Substituting the expansion (\ref{FadCompSplinExp}) into  Eq.
(\ref{EqFaddTAM1}) followed by the orthogonal collocation procedure one gets
a system of linear algebraic equations
\begin{equation}
\label{Deq}
	(\hat{H}^0_0+\hat{V}(\hat{I}+\hat{P})-E\hat{I}){\bf f}=0 \: .
\end{equation}
where $\hat{H}^0_0 $,  $ \hat{V} $,  $ \hat{I} $, 
and  $ \hat{P}  $ are the 
matrices corresponding to the 
operators of  Eq. (\ref{EqFaddTAM1}) while 
$ \bf f$ is the coefficient vector  of the 
expansion (\ref{FadCompSplinExp}).

The dimension of this system is rather large and equals  
$ N=N^{c}_{x}N^{c}_{y}N^{c}_{z} $. Therefore, we  use this system 
of equations to  find only its lowest generalized eigenvalues 
and eigenvectors.
\section{The Tensor-trick for Cartesian Coordinates}
The inversion  procedure of the operator  $(\hat H_1-E\hat I)$ is
crucial for the performance of the algorithm and therefore a brief 
description of a tensor-trick algorithm  for the DFE 
in Cartesian coordinates  will be given. Furthermore, as the matrices 
involved are large, certain comments will be passed concerning
the efficiency in computation.\\

Consider the Hamiltonian $\hat H_1=\hat H^0_0+\hat V\hat I$. 
 The structure of this operator is
\begin{eqnarray*}
	\widehat I&=&\widehat I_x\otimes \widehat I_y\otimes \widehat I_z\\
             		\widehat H_1&=&\widehat D_x\otimes \widehat I_y
			\otimes \widehat I_z
			+\widehat I_x
			\otimes \widehat D_y\otimes \widehat I_z\\
                      &+&\widehat I_x
			\otimes \widehat I_y\otimes \widehat 1_z\,(\widehat L_x
			\otimes \widehat 1_y+\widehat 1_x\otimes \widehat L_y)
			\otimes \widehat D_z
			 +\widehat V\widehat I_x\otimes 
                        \widehat I_y\otimes \widehat I_z\, .
\end{eqnarray*}
The  $\hat I_t$ are the matrices corresponding to the identity 
operators acting in coordinate $t$,  
$\hat 1_t $ is the  unit matrix,
and  $ \hat D_t $ contains the elements of the corresponding 
differential operators. The tensor structure of the operator  
$ \hat H $ allows a  diagonalisation of it without solving the 
spectral problem in the  $ N $--dimensional space.
To perform this diagonalisation one introduces the 
following matrices  $ \widehat B_z$,  $ \widehat B_x^i$,  
$ \widehat B^i_y$  constructed from  the generalized
eigenvectors of the operators acting in different coordinates,
\begin{eqnarray*}
	 \widehat B_z\ &:&\  \widehat D_z\widehat B_z \,=\,\widehat L_z
                  \widehat I_z\widehat B_z\\
	\widehat B^i_x \ &: &\ (\widehat D_x+\widehat V\widehat I_x+
		l_i\widehat L_x\widehat I_x)\widehat B^i_x\,=\,\widehat X^i
		\widehat I_x\widehat B^i_x\\
	\widehat B^i_y\ &: &\ (\widehat D_y+l_i\widehat L_y
	\widehat I_y)\widehat B^i_y\,=\,\widehat Y^i\widehat I_y
	\widehat B^i_y
\end{eqnarray*}
where  
	$ \hat L_z={\rm diag} \{l_1,l_2,\ldots ,l_{Nz}\} $,  
	$\widehat X^i={\rm diag } \{E^i_{x1},E^i_{x2},
	\ldots ,E^i_{xN_x}\} $,
and 	 $ \widehat Y^i\,=\,{\rm diag} \{E^i_{y1},E^i_{y2},
	\ldots ,E^i_{yN_y}\} $, and $E^i_{tN_t}$ are the eigenvalues
of the corresponding operators.
One also defines the matrices  
	$\widehat B_z^*$,  
	$\widehat B_{x^i}^*$,
 	$\widehat B_{i_y}^*$  
constructed of the eigenvectors of the adjoined
equations. Expanding the expression  
$\hat B^*(\hat H_1-E\hat I)\hat B$,
with   $\hat B $ and  $\hat B^*$ being given by
\begin{eqnarray*}
	\widehat B &\equiv &\prod ^{N_z}_{i=1}\oplus 
	(\widehat B^i_x\otimes \widehat B^i_y)
	\otimes \widehat{[B}_z]_i\\
	\widehat B^*&\equiv& \prod^{N_z}_{i=1}\oplus 
	(\widehat {B^i}^*_x\otimes \widehat {B^i}^*_y)
	\otimes \widehat{[B}^*_z]_i\,,
\end{eqnarray*}
and taking into account the relation  
$ \hat B^*\hat B=\hat 1_x \otimes \hat 1_y\otimes \hat 1_z$,
one can prove that the matrices  $\hat B$ and  $\hat B^*$ are the
desired matrices diagonalising the operator  $ (\hat H_1-E\hat I) $. 
Once this operator is diagonalised it can then be easily inverted
\begin{eqnarray*}
	&& \widehat B^*(\hat H_1-E\hat I)\widehat B=\widehat G\,,\\
	&&(\hat H_1-E\hat I)^{-1}=\widehat B\widehat G^{-1}
       \widehat B^*\, \\
&&	\widehat G={\rm diag}\{ g_{111},g_{112},\ldots,
	g_{i_xi_yi_z},\ldots ,g_{N_x N_y N_z}\}\\
	&&  g_{i_x i_y i_z}=E^{i_z}_{xi_x}+E^{i_z}_{xi_y}-E
\end{eqnarray*}
Thus, in principle, the task of construction of the operator 
$ (\hat H_1-E\hat I)^{-1}$ is over. However, to use computer
power effectively one must care about optimal implementation of
all the operations involved in the computation of the product of
the operator  $ \hat h^{-1}(E)\equiv (\hat H_1-E\hat I)^{-1}$ and a vector  
${\bf u}$, the number of arithmetic operations should be minimized. To
implement the operation effectively the structure of the operator
 $ \hat h^{-1}(E) $ can be used 
$$
	\hat h^{-1}(E) \equiv \widehat B
  		\widehat G^{-1}(E)\widehat B^*\, .
$$
The matrix  $\widehat G^{-1}(E)$  has explicit diagonal structure, and the
calculation of the product $ \widehat G^{-1}{\bf  u}$ requires only 
$O(N_x N_y N_z)$ 
 operations. Since however, the matrices $\widehat B$ and
$\widehat{B}^{*}$ are dense, and in principle the calculation of the
products $\widehat B^{(*)}{\bf u}$ requires $O((N_x N_y N_z)^2)$ arithmetic
operations, that could essentially decrease the effectiveness of the method.
However, taking into account the structure of the matrices  $\widehat{B}$ 
and  $\widehat{B}^{*} $ one can improve the performance. Consider the 
procedure of tensor multiplication of two matrices by a vector ${\bf u}$:
$$
	{\bf v}=\widehat{B_{x}}\otimes \widehat{B_{y}}{\bf u}\, ,
$$
and use the identity
$$
	\widehat {B_x}\otimes \widehat{B_y}=\widehat I\otimes 
	\widehat{B_y}\widehat{B_x}\otimes \widehat I\,.
$$
Obviously, even the calculation of the product 
$ \widehat{B_x}\otimes \widehat{B_y}{\bf u}$  
requires $N^2_xN^2_y$ operations, the calculation of the products
$\widehat I\otimes \widehat{B_y}{\bf u}$ and  $ \widehat{B_x}\otimes \widehat I{\bf u}$  
require correspondingly only  $N^{2}_{x}N_{y}$ and  $N_{x}N^{2}_{y}$ operations.
Therefore a multiplication of a tensor product by a vector 
would cost $ N_x N_y (N_x +N_y)$ operations.
Similar relations can be written for a direct sum. 
Thus the correct implementation of a direct sum multiplication by a vector 
can reduce the computation costs  $N_x N_y N_z/(N_x+N_y+N_z)$  
times.

The computation costs can be reduced even further. One of the 
possibilities is to abandon full diagonalisation of the operator
$ \widehat h(z)  $ and to modify the computation scheme.
For this, let us construct the matrices diagonalising only the terms 
corresponding to the coordinates  $y$ and $z$ 
\begin{eqnarray*}
&&	\tilde B\equiv \prod^{N_z}_{i=1}
	           \oplus (\widehat I \otimes \widehat{B^i}_y)
                           \otimes \widehat{[B_z]}_i\, ,\\
&&
	\tilde B^*\equiv \prod^{N_z}_{i=1}
	             \oplus (\widehat I \otimes \widehat{B^i}^*_y)
                            \otimes \widehat{[B^*_z]}_i\, .
\end{eqnarray*}
In this case the matrix  
$ \tilde G(z)\equiv \tilde B^{*} \widehat h(z)\tilde B$  
is band because of the fact that the operator is local and the basic functions
have well localized support (\ref{SplineBases}).
The multiplication $ \widehat{L}(z){\bf u}$ can be performed in two stages:
\begin{itemize}
\item[i)] 
Solve a system of equations with a 
band matrix $\tilde{G}(E) $  
$$
	\tilde{G} (z)\tilde{{\bf u}}=\tilde B^*
	(\widehat V \widehat P+\widehat V_c\widehat S){\bf u}\, ;
$$
\item[ii)] 
calculate  $ {\bf v}\equiv \widehat{L}(z){\bf u}$ as
$$
	{\bf v}=\tilde B\tilde{{\bf u}}\, .
$$
\end{itemize}
Taking into account the tensor structure of the matrices 
$\tilde B^*$ and $\tilde B$ this computation scheme requires only 
$ O(N_x\,N_y\,N_z\,(N_y+N_z))$ operations.
%
\section{ Convergence rates for the Arnoldi algorithm}
The key point of this algorithm  is the repeated action of the 
operator under consideration on a sequence of vectors generating 
a Krylov subspace where the desired eigenvector
lies. To use this algorithm to solve the eigenvalue problem (\ref{Lambda1})
one must be able to calculate the action of the operators  
$(\hat H_1-E\hat I)^{-1} $ and  $ \hat H_2$ on an arbitrary vector  ${\bf u}$. 
Thus the problems of inversion of the operator  
$ (\hat H_1-E\hat I )  $ and storage of the matrix
 $ \hat H_2 $ should be solved first. In Ref. \cite{Groning} hyperspherical
coordinates were used which lead to an effective storage scheme for the 
matrices 
 $ \hat H_2=\hat V(\hat I+\hat P)$. However, one
can not include pair interaction into the invertible part  
$ (\hat H_1-E\hat I) $   using hyperspherical coordinates. 
On the other hand, if  Cartesian coordinates are used then one has  
difficulties with the storage of the matrix  $ \hat P$  since these 
coordinates do not have the same symmetry properties
as the operator  $P$ and this results  to an irregular
structure of $ \hat P$. 
In our case the situation is saved by the  total angular momentum 
representation. In contrast to bispherical expansion used 
in \cite{Groning}, the operator  $P$ is local and does not 
contain any terms requiring integration. As a result the matrix  
$ \hat P$ is sparse and can be either effectively
stored using any standard storage scheme for sparse matrices 
\cite{sparce} or not stored at all. In the last case the matrix 
elements are calculated ``on fly'' when required. 

The estimations of the convergence rate of the  Arnoldi algorithm \cite{Saad} 
are essential in understanding the problems that arise when the
tensor-trick algorithm is applied. Suppose $\hat{L}$ is a linear 
operator in a finite-dimensional space with
$$
	 \hat L {\bf u}_i= \lambda_i {\bf u}_i \, .
$$
Let ${\cal K}_m(\bf v)$  be a Krylov subspace of dimension $m$ constructed 
from a  starting vector $\bf v$ of the operator $\hat L $ and ${\cal P}_m$ be 
an orthogonal projector to ${\cal K}_m$ space. The
convergence rate of the algorithm can be expressed in terms of the norm
$$
	\varepsilon_i^{(m)} \equiv ||(1-{\cal P}_m) {\bf u}_i||
$$
of the eigenvector $u_i$ projection onto the orthogonal
completion to ${\cal K}_m$. Suppose $\alpha_k$ are the expansion 
coefficients of the
starting vector ${\bf v}$ 
in terms of the eigenvectors
${\bf u}_i$,
$$
	  {\bf v}=\sum_{i=1}^{n} \alpha_i {\bf u}_i \, .
$$
Then the following relation is fulfilled
$$
 \varepsilon_i^{(m)} \le \xi_i \epsilon_i^{(m)} \, ,
$$
where $\xi_i = -1+\sum_{k=1}^n |\alpha_k|/|\alpha_i |$
and $\epsilon_i^{(m)}$ depends on the spectral properties of the operator
$\hat L $. For the eigenvalues lying in the external part of the spectrum some
practically important estimations are known. Suppose that all the eigenvalues of
$\hat L $ but $\lambda_1$ lye inside of an ellipse with the center $c$, largest
semi-axis $a$, and focuses $c \pm e$.  
It can then be shown that  $\epsilon_1^{(m)}$ satisfy the 
following estimation \cite{Saad}
$$
 \displaystyle 
  |\epsilon_1^{(m)}| \le \frac{C_{m-1}(\displaystyle
  \frac{a}{e})}{C_{m-1}(\displaystyle \frac{\lambda_1-c}{e})} \ ,
$$
where $C_{m}(t)$ are Chebyshev polynomials of the first kind. 

Consider now the convergence rate of Arnoldi algorithm for the operator $\hat L (z)$
(\ref{Lz}) for the value of z corresponding to the ground state of a three-body
system. In this case the maximal eigenvalue $\lambda_1=1$. Having defined the
separation distance between the largest eigenvalue and the rest part of the
spectrum $s \equiv |1- \lambda_2|$, the distance between the maximal eigenvalues
in the rest part $d \equiv |\lambda_2 -\lambda_n |$ and neglecting the imaginary
part of the spectrum,  we obtain the estimation
$$
  |\epsilon_1^{(m)}| \le |C_{m-1}(1+\frac{2s}{d})|^{-1} \ .
$$
Therefore the parameter managing the convergence rate is the ratio $s/d$
of the separation to the size of the rest part of the spectrum. The convergence
rate could be improved by enlarging the separation and by enhancing
the remainder part of the spectrum localization.
%

\begin{table}
\begin{center}
\caption{Bound state energies for the helium dimer He$_2$ and trimer He$_3$: 
$ E_2^{\rm ex} $ and $E_2$  are the exact and  calculated  
(using the same grid employed in three-body calculations)  
two-body bound states while  $E_3$ and $E_3^*$ are the 
 ground and excited  states for the trimer. \label{tabTrimerEv} }
\vspace{3mm}
\begin{tabular}[c]{l c c c c }
Potential & 
$E_2^{\rm ex}$ (mK)  &  $E_2$ (mK)  &  $E_3$ (K)  &  $E^*_3$ (mK) \\
\hline 
 SAPT1\cite{Az97}    	&  -1.732405  &  -1.7322  &  -0.13382  &  -2.790 \\
 SAPT2\cite{Az97}    	&  -1.815003  &  -1.8146  &  -0.13516  &  -2.888 \\
 SAPT \cite{Az97}    	&  -1.898390  &  -1.8979  &  -0.13637  &  -2.986 \\
 LM2M2\cite{Az91}   	&  -1.303482  &  -1.304   &  -0.12641  &  -2.271 \\
 TTYPT\cite{TTY}    	&  -1.312262  &  -1.3121  &  -0.12640  &  -2.280 \\ 
 HFD-B3-FC11\cite{Az97} &  -1.587301  &  -1.5873  &  -0.13126  &  -2.617 \\
\end{tabular}
\end{center}
\end{table}  
\vskip 1cm
\begin{table}[p]
\begin{center}
\caption{Convergence for the He$_3$ ground state energy with respect 
to the number of grid points for the LM2M2 potential. 
\label{tabConv2}  }
\vspace{5mm} 
\begin{tabular}[c]{l c c }
Grid                    & $E_3$ (K) & $E_2$ (mK)\\
\hline  
  45$\times $45$\times $6   & -0.12570 & -1.3002 \\
  60$\times $60$\times $6   & -0.12583 & -1.3032\\
  75$\times $75$\times $6   & -0.12582 & -1.3034 \\
  90$\times $90$\times $6   & -0.12582 & -1.3035 \\
105$\times $105$\times $6   & -0.12581 & -1.3034 \\
105$\times $105$\times $9   & -0.12636 & -1.3034 \\
105$\times $105$\times $12  & -0.12636 & -1.3034 \\
105$\times $105$\times $15  & -0.12640 & -1.3034 \\
105$\times $105$\times $18  & -0.12640 & -1.3034 \\
\end{tabular} 
\end{center}
\end{table}  
%
\begin{table}[p] 
\begin{center}
\caption{Same as in Table \ref{tabConv2} but for the  excited state.
\label{tabConv} }
\vspace{5mm} 
\begin{tabular}[c]{ l c c} 
Grid & $E^*_3$ (mK) & $E_{2}$ (mK)\\
\hline 
  45$\times $45$\times $6   & -2.2658 & -1.3011 \\
  60$\times $60$\times $6   & -2.2649 & -1.3018 \\
  75$\times $75$\times $6   & -2.2668 & -1.3031 \\
  90$\times $90$\times $6   & -2.2670 & -1.3031 \\
105$\times $105$\times $6   & -2.2677 & -1.3037 \\
105$\times $105$\times $9   & -2.2712 & -1.3037 \\
105$\times $105$\times $12  & -2.2707 & -1.3037 \\
105$\times $105$\times $15  & -2.2707 & -1.3037 \\
\end{tabular}
\end{center}
\end{table} 
\begin{table}
\begin{center}
\caption{Comparison of the results obtained with  the LM2M2 potential 
with other  results in the field.
\label{tabCompare}
}
\vspace{5mm}
\begin{tabular}[c]{lcccc}
 Observable 		& This work    & \cite{Fedorov}	& \cite{Moto2} \\
 \hline
 $E_{3}$, K               & -0.1264      & -0.1252 	& -0.1259     \\
 $E^*_{3}$, mK            & -2.271       & -2.269  	& -2.28       \\
 $<r^{2}>^{1/2}$, \AA     &  6.32        &  6.24   	& 	      \\
 $<r^{2}_{*}>^{1/2}$, \AA &  59.3        &  60.86  	& 	      \\ 
\end{tabular}
\end{center}
\end{table} 
\begin{table}
\begin{center}
\caption{The mean radius of He$_{3}$ in  \AA 
\label{tabR} }
\vspace{5mm}
\begin{tabular}[c]{lccc}
Potential & Ground state of He$_{3}$ & Excited state of He$_{3}$ & He$_{2}$\\
\hline 
 SAPT1         & 5.47 &  47.5  & 45.60 \\
 SAPT2         & 5.45 &  47.1  & 44.63 \\
 SAPT          & 5.44 &  47.1  & 43.73 \\
 HFD-B3-FC11   & 5.49 &  48.3  & 47.47 \\
 LM2M2         & 5.55 &  49.9  & 52.00 \\
 TTYPT         & 5.55 &  50.1  & 51.84 \\ 
\end{tabular}
\end{center}
\end{table} 
\begin{table}
\caption{The mean square radius of He$_{3}$ in \AA 
\label{tabR2} }
\begin{center}
\vspace{5mm}
\begin{tabular}[c]{lccc}
Potential & Ground state  & Excited state  & He$_{2}$\\
\hline 
 SAPT1         & 6.22 & 56.4 & 61.89 \\
 SAPT2         & 6.21 & 55.9 & 60.52 \\
 SAPT          & 6.19 & 55.9 & 59.24 \\
 HFD-B3-FC11   & 6.26 & 57.4 & 64.53 \\
 LM2M2         & 6.32 & 59.3 & 70.93 \\
 TTYPT         & 6.33 & 59.6 & 70.70 \\ 
\end{tabular}
\end{center}
\end{table} 
\begin{table}
\begin{center}
\caption{Contribution of cluster wave to the Faddeev component
\label{tabClusterisat}
        }
\vspace{5mm}
\begin{tabular}[c]{ l c c }
Potential & $\Vert f_c\Vert^2$ & $\Vert f_c^*\Vert^2$\\
\hline 
 SAPT1         & 0.2743 & 0.9442 \\
 SAPT2         & 0.2787 & 0.9461 \\
 SAPT          & 0.2832 & 0.9478 \\
 HFD-B3-FC11   & 0.2661 & 0.9407 \\
 LM2M2         & 0.2479 & 0.9319 \\
 TTYPT         & 0.2487 & 0.9323 \\
\end{tabular}
\end{center}
\end{table} 
\begin{table}
\newpage
\begin{center}
\caption{Contribution of different two-body angular states to the Faddeev component
\label{tabAngContribFC} 
}
\vspace{5mm}
\begin{tabular}[c]{lllllll}
 & 
\multicolumn{3}{l}{Ground state} & 
\multicolumn{3}{l}{Excited state}\\
\hline Potential & S & D & G & S & D & G\\
\hline 
 SAPT1         & 0.9989914 & 0.0009974 & 0.0000108 & 0.9999951  & 0.0000049 & 0.0000001 \\
 SAPT2         & 0.9989822 & 0.0010065 & 0.0000109 & 0.9999950  & 0.0000050 & 0.0000001 \\
 SAPT          & 0.9989754 & 0.0010131 & 0.0000111 & 0.9999949  & 0.0000050 & 0.0000001 \\
 HFD-B3-FC11   & 0.9990060 & 0.0009830 & 0.0000106 & 0.9999953  & 0.0000047 & 0.0000000 \\
 LM2M2         & 0.9990393 & 0.0009500 & 0.0000103 & 0.9999956  & 0.0000043 & 0.0000000 \\
 TTYPT         & 0.9990332 & 0.0009561 & 0.0000104 & 0.9999956  & 0.0000043 & 0.0000000 
\end{tabular}
\end{center}
\end{table} 
\begin{table}
\begin{center}
\caption{Contribution of different two-body angular states to the wave function
\label{tabAngContribWF}
}
\vspace{5mm}
\begin{tabular}[c]{lllllll}
 & \multicolumn{3}{l}{Ground state \ } & \multicolumn{3}{l}{ \ Excited state}\\
\hline Potential & S & D & G       & \  S & D & G\\
\hline 
 SAPT1         & 0.9521 & 0.0336 & 0.0089\ & \ 0.855 & 0.099 & 0.031 \\
 SAPT2         & 0.9520 & 0.0338 & 0.0090 & \ 0.854 & 0.100 & 0.031 \\
 SAPT          & 0.9519 & 0.0339 & 0.0090 & \ 0.826 & 0.102 & 0.034 \\ 
 HFD-B3-FC11   & 0.9525 & 0.0335 & 0.0089 & \ 0.837 & 0.100 & 0.033 \\
 LM2M2         & 0.9530 & 0.0329 & 0.0089 & \ 0.873 & 0.093 & 0.025 \\
 TTYPT         & 0.9527 & 0.0332 & 0.0094 & \ 0.862 & 0.094 & 0.029 \\ 
\end{tabular}
\end{center}
\end{table} 

\end{document}